\documentclass[twocolumn,floats,prb,aps,showpacs]{revtex4}
\usepackage{graphics,epsfig,graphicx}
\usepackage{color}

\usepackage{latexsym}
\usepackage{calrsfs}
\usepackage{amsmath}
\usepackage{graphicx}
\usepackage{amsfonts}
\usepackage{amsmath}
\usepackage{amssymb}
\usepackage{exscale}
\usepackage{color}

\begin{document}



\title{Optical signatures of a fully dark exciton condensate}

\author{Monique Combescot$^1$, Roland Combescot$^{2,3}$, Mathieu Alloing$^4$, Fran\c{c}ois Dubin$^4$}

\affiliation{
$^{1}$ Institut des Nanosciences de Paris, Universit\'{e} Pierre et Marie Curie, CNRS, 2 pl. Jussieu,
75005 Paris, France}

\affiliation{
$^{2}$ Laboratoire de Physique Statistique, Ecole Normale Sup\'{e}rieure, Universit\'{e} Pierre et Marie Curie, 24 rue Lhomond, 75005 Paris,
France}

\affiliation{
$^{3}$ Institut Universitaire de France, 103 boulevard Saint-Michel, 75005
Paris, France}

\affiliation{
$^{4}$ ICFO-The Institute of Photonic Sciences, 3
Av. Carl Friedrich Gauss, 08860
Castelldefels (Barcelona), Spain}

\date{\today }

\begin{abstract}
We propose optical means to reveal the presence of a dark exciton condensate that does not yield any photoluminescence at all. We show that (i) the dark exciton density can be obtained from the blueshift of the excitonic absorption line induced by dark excitons; (ii) the polarization of the dark condensate can be deduced from the blueshift dependence on probe photon polarization and also from Faraday effect, linearly polarized dark excitons leaving unaffected the polarization plane of an unabsorbed
photon beam. These effects result from carrier exchanges between dark and bright excitons.
\end{abstract}

\maketitle


The experimental observation of exciton Bose-Einstein condensation faced 
major difficulty for decades\cite{Snoke_Book}. Experiments
mostly looked for signatures in an intense photoluminescence emitted by 
bright excitons with total spin ($\pm$1), thus overlooking that optically inactive states, i.e, dark states with total spin ($\pm$2), have a lower energy\cite{SSC_MC_ML}. 
This fact was known already\cite{Blackwood_94,Ekardt_79,Vina_99,Amand_97} 
but its consequence on exciton Bose-Einstein condensation 
was put forward very recently only. We showed\cite{PRL_MC_OBM_RC}
that, in the dilute regime, the exciton Bose-Einstein condensate must be purely dark. So,
such a condensate cannot be detected through a photoluminescence signal. However, above a density
threshold, the
condensate acquires a bright coherent component\cite{PRL_RC_MC} through carrier exchanges which couple dark and bright excitons: it becomes "gray" and can be detected through the
weak photoluminescence emitted by its bright part.

In very recent experiments\cite{us_arxiv1}, we have revealed such a "gray" condensate in the vicinity of a fragmented ring-shaped exciton gas\cite{us_arxiv1,us_arxiv2,butov_nature2002,butov_nature2012}. Precisely, we have
shown that  (i) a far denser dark exciton gas coexists with bright
excitons, (ii) the bright component exhibits a coherence length 10 times larger
than  the thermal de Broglie wavelength and a linear polarization expected for
degenerate ($\pm 1$) states, (iii) these two features disappear when the bath
temperature gets larger than a few Kelvins\cite{us_arxiv1}.

Revealing a fully dark exciton condensate, 
at density smaller than the threshold for appearance of
a bright component, remains a real challenge. This would be really valuable because
we would then have a direct evidence for the true dark nature of the condensate. A dark
condensate can be already guessed in a trap\cite{SSC_MC_ML} confining both, bright and
dark excitons, e.g.,  
electrostatic traps\cite{Rapaport_05,Gartner_07,High_09,Alloing_13}, traps induced by stress\cite{Sinclair_11} or even optical traps\cite{PRL_MC_MM_CP}. Indeed, as the bath temperature is decreased,
quantum condensation must lead to a darkening of the trap center,
where the potential energy is minimum. 

In this letter, we propose an all-optical approach to reach both, the density
and the spin polarization of a fully dark exciton condensate\cite{PRL_RC_MC}. We show that
the condensate density can be quantitatively deduced from 
the absorption spectrum of a probe beam: carrier exchanges
between a dark exciton gas and a bright exciton produced by a probe photon yield a blueshift
of the excitonic absorption line which reveals the dark exciton density. 

To reach the spin polarization of the dark condensate, we must use probe photons tuned on an excited exciton level. One can also use Faraday effect: for linearly polarized dark excitons, the polarization plane of a linearly polarized unabsorbed beam stays unaffected while for different dark exciton polarization, a Faraday rotation would be observed. This is in contrast with similar effects involving bright excitons: In the exciton optical Stark effect, the blueshift of a probe beam strongly depends on pump and probe polarizations even for probe photons tuned on the ground state, while a linearly polarized bright exciton gas induces a "Faraday oscillation", the polarization changing from linear to elliptical and back to linear as photons propagate through the sample\cite{SSC_MC_OBM_2009}.

\begin{figure*}\label{fig1}
\includegraphics[width=18cm]{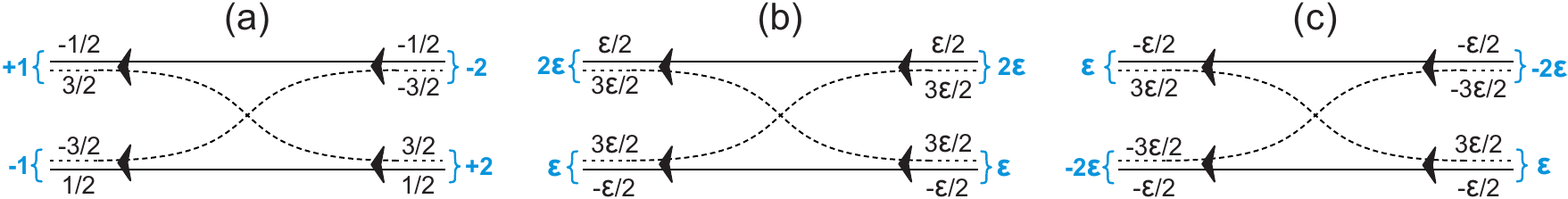}
\caption{(a) : Hole exchange between two excitons with opposite polarization. (b) and (c) : hole exchange between a dark and a bright exciton ($\epsilon=\pm1$).}
\end{figure*}

{\bf I. Absorption line shift induced by dark excitons}. Dani\`{e}le Hulin and coworkers\cite{D_Hulin}  have discovered that the absorption line of a probe beam blueshifts when the semiconductor is irradiated by an unabsorbed pump beam. This shift results from carrier exchanges between the real (bright) exciton created by a probe photon and virtual (bright) excitons coupled to the unabsorbed pump beam. We here consider a similar shift resulting from exchange between a real (bright) exciton created by a probe photon and dark excitons possibly present in the sample due to Bose-Einstein condensation. To get this shift, we follow the procedure detailed in Ref.\cite{Monique_Stark}.

Without dark excitons, the initial state before photon absorption is the electron-hole pair vacuum $| 0 \rangle$ with energy 0. The final state after the absorption of a probe photon with momentum $\textbf{Q}_{p}$ tuned on the exciton level $\nu_{p}$,  is the excitonic state $B^{\dagger}_{p} | 0 \rangle$ with
\begin{equation}\label{eq1}
    B^{\dagger}_{p}=a_{{+}1}B^{\dagger}_{\nu_{p},\textbf{Q}_{p};+1}+a_{{-}1}B^{\dagger}_{\nu_{p},\textbf{Q}_{p};-1},
\end{equation}
its energy being $E_{p}=\epsilon_{\nu_{p}}+Q^{2}_{p}/2(m_{e}+m_{h})$, where $m_e$ and $m_h$ are the electron and hole masses and $\epsilon_{\nu_{p}}$ the exciton relative motion energy while $a_{\pm1}$ are the amplitudes of the photon $\sigma_{\pm}$ polarization. The creation operator $B^{\dagger}_{i;\pm1}$ for a bright exciton with total spin $S_i=\pm1$ in state $i \equiv(\nu_{i},\textbf{Q}_i)$, is related, in narrow quantum wells, to free electron-hole creation operators through
\begin{equation}\label{eq2}
    B^{\dagger}_{i;\pm1}=\sum_{\textbf{p}}a^{\dagger}_{\textbf{p}+\gamma_{e}\textbf{Q}_{i};\mp1/2}b^{\dagger}_{-\textbf{p}+\gamma_{h}\textbf{Q}_{i};\pm3/2} \langle\textbf{ p} | \nu_{i}\rangle
\end{equation}
 where $\gamma_{e}=1-\gamma_{h}=m_{e}/(m_{e}+m_{h})$. When the probe photons have an elliptical polarization $\phi$ with main axes $(\textbf{X},\textbf{Y})$ tilted by $\theta$ from the $(\textbf{x},\textbf{y})$ well axes, the prefactors in Eq.\ref{eq1} are $a_{+1}=\mathrm{e}^{-i\theta}\cos(\phi-\pi/4)$ and $a_{-1}=\mathrm{e}^{i\theta}\sin(\phi-\pi/4)$. Photons linearly polarized along $\textbf{x}$ correspond to ($\theta =0,\phi =0$) while $\sigma _{+}$ photons correspond to ($\phi =\pi/4)$.

Let us write as $|\Psi_{N}\rangle$ the initial state in the presence of a dark exciton condensate, its energy being $\mathcal{E}_{N}$. In the dilute limit, $|\Psi_{N}\rangle$ is close to
\begin{equation}\label{eq3}
|\Psi_{N}\rangle\simeq (D^{\dagger})^{N}|0\rangle=(a_{+2}B^{\dagger}_{\nu_{0},\textbf{0};+2}+a_{-2}B^{\dagger}_{\nu_{0},\textbf{0};-2})^{N}|0\rangle.
\end{equation}
Creation operators for dark excitons $S_i=\pm2$ are related to electron-hole creation operators through
\begin{equation*}
    B^{\dagger}_{i;\pm2}=\sum_{\textbf{p}}a^{\dagger}_{\textbf{p}+\gamma_{e}\textbf{Q}_{i};\pm1/2}b^{\dagger}_{-\textbf{p}+\gamma_{h}\textbf{Q}_{i};\pm3/2}\langle \textbf{p}|\nu_{i}\rangle,
\end{equation*}
the dark-bright exciton splitting having negligible effects on the exciton relative motion.

The final state $|\Phi_{N+1}\rangle$ after photon absorption is close to $B^{\dagger}_{p}|\Psi_{N}\rangle$.
Let $|X_{N,p}\rangle$ 
be its normalized form. To get $|\Phi_{N+1}\rangle$, we introduce the projector $P_{\bot}$ over the subspace orthogonal to $|X_{N,p}\rangle$
defined through
\begin{equation}\label{eq4}
I=|X_{N,p}\rangle \langle X_{N,p}|+P_{\bot}.
\end{equation}
 Inserting this identity in front of $|\Phi_{N+1}\rangle$ in $(H_{sc}-\mathbf{E}_{N+1})|\Phi_{N+1}\rangle=0$ where $H_{sc}$ is the semiconductor hamiltonian and multiplying the resulting equation by $P_{\bot}$ yields

\vspace{-15 pt}

\begin{multline}\label{eq5}
   P_{\bot}|\Phi_{N+1}\rangle= \\
   \Big[P_{\bot} (\mathbf{E}_{N+1}-H_{sc}) P_{\bot}\Big]^{-1}P_{\bot}H_{sc}|X_{N,p}\rangle\langle X_{N,p}|\Phi_{N+1}\rangle.
\end{multline}
When used into the Schr\"{o}dinger equation projected onto $\langle X_{N,p}|$, we end with 
\begin{multline}\label{eq6}
\mathbf{E}_{N+1}=\langle X_{N,p}|H_{sc}|X_{N,p}\rangle+\\\langle X_{N,p}|H_{sc}P_{\bot}\Big[P_{\bot} (\mathbf{E}_{N+1}-H_{sc}) P_{\bot}\Big]^{-1}P_{\bot}H_{sc}|X_{N,p}\rangle.
\end{multline}
The shift of the probe photon absorption line induced by the dark exciton condensate is given by

\vspace{-10 pt}

\begin{equation}\label{eq7}
\Delta = (\mathbf{E}_{N+1}-\mathcal{E}_{N})-E_{p}.
\end{equation}

\noindent A simple way to get it from Eq.(\ref{eq6}) is to introduce the "creation potential" $V_{p}^{\dagger}$ for the $B_{p}^{\dagger}$ exciton defined as
\begin{equation}\label{eq8}
\left[ H_{sc},B_{p}^{\dagger}\right]=E_{p}B_{p}^{\dagger}+V_{p}^{\dagger}.
\end{equation}
Since $(H_{sc}-\mathcal{E}_{N})|\Psi_{N}\rangle=0$, we then find
\begin{equation}\label{eq9}
H_{sc}B_{p}^{\dagger}|\Psi_{N}\rangle=(E_{p}+\mathcal{E}_{N})B_{p}^{\dagger}|\Psi_{N}\rangle+V_{p}^{\dagger}|\Psi_{N}\rangle.
\end{equation}
So, for $H_{sc}$ acting either on the right or on the left in the first term of Eq.(\ref{eq6}), we get
\begin{multline}\label{eq10}
\Delta = \frac{1}{2}\frac{\langle\Psi_{N}|B_{p}V_{p}^{\dagger}+V_{p}B_{p}^{\dagger}|\Psi_{N}\rangle} {\langle\Psi_{N}|B_{p}B_{p}^{\dagger}|\Psi_{N}\rangle} +\\\frac{\langle\Psi_{N}|V_{p}\Big[P_{\bot}(\Delta+E_{p}+\mathcal{E}_{N}-H_{sc})P_{\bot}\Big]^{-1}V_{p}|\Psi_{N}\rangle} {\langle\Psi_{N}|B_{p}B_{p}^{\dagger}|\Psi_{N}\rangle}.
\end{multline}
The first term comes from one Coulomb interaction between the probe exciton and the condensate while the second term corresponds to correlations.

\begin{figure*}\label{fig2}
\includegraphics[width=18cm]{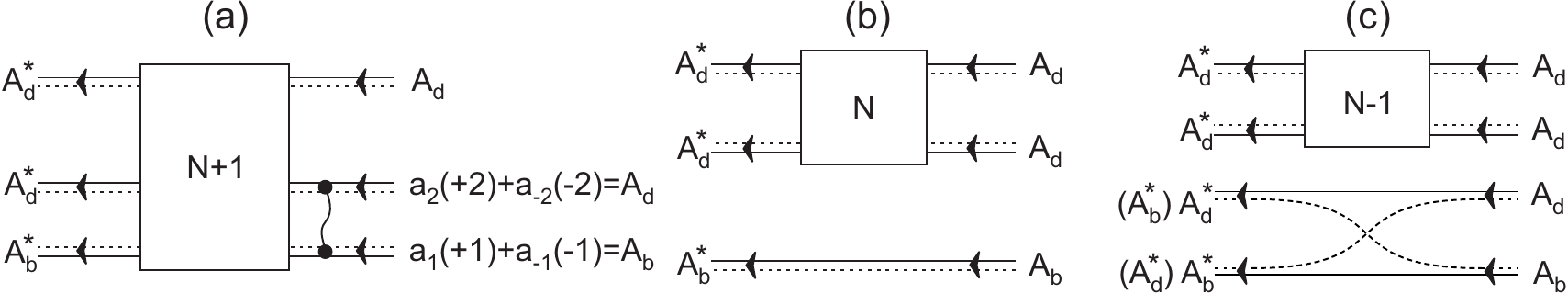}
\caption{(a): Shiva diagram for the scalar product $\langle \Psi_{N}| B_{p}V_{p}^{\dagger} | \Psi_{N}\rangle$. 
(b and c): zero and first order terms in dark exciton density.}
\end{figure*}

We can calculate these terms by using the many-body formalism for composite bosons we have recently developed \cite{Phys_Rep}. Its form with spin degrees of freedom can be found in Ref. \cite{PRB_MC_OBM_2006}. However, a
physically more enlightening way to perform many-body calculations is to use Shiva diagrams. Readers not
familiar with these diagrams can recover the following results using the relevant set of commutators recalled in the Supplementary Material.

The denominator in Eq.(\ref{eq10}) is represented by Fig.2a without the wavy coulomb process. Its first term, shown in Fig.2b, is given by
\begin{equation}\label{eq17}
    \langle \Psi_{N} | \Psi_{N}\rangle\simeq \langle 0 | D^{N}(D^{\dagger})^{N}|0\rangle=N!F_{N}.
\end{equation}
 $F_{N}$ comes from carrier exchanges. For large $N$, it is exponentially small, but
 $F_{N-1}/F_{N}=1+\mathcal{O}(\eta)$ where $\eta=N(a_{X}/L)^{d}$ is the dimensionless gas parameter ruling many-body effects between $N$ excitons in $d$ dimension. The second term, shown in Fig.~2c, contains a factor $N^2$ from the $N$ ways, on each side, to choose the dark exciton involved in carrier exchange. Contributions of each exchange process shown in Figs.(1b,1c) add up, leading to a $\Sigma_{\epsilon}\left[\lambda(_{p\;\;p}^{0\;\;0})|a_{\epsilon }|^2|a_{2\epsilon }|^2
+\lambda(_{0\;\;p}^{p\;\;0})|a_{\epsilon }|^2|a_{-2\epsilon }|^2\right]$ factor. So, this second term gives
\begin{multline}\label{eq18}
        -N^{2}\bigg[\lambda(_{p\;\;p}^{0\;\;0})
  \left( \left| a_{1}a_{2} \right|^{2}+\left| a_{-1}a_{-2} \right|^{2}\right)\\ 
  +\lambda(_{p\;\;0}^{0\;\;p})\left(\left| a_{1}a_{-2} \right|^{2}+\left| a_{-1}a_{2} \right|^{2} \right)\bigg](N-1)!F_{N-1}.
\end{multline}
As $\lambda(_{0\;\;0}^{0\;\;0}){=}\lambda_{d}(a_{X}/L)^{d}$ with $\lambda_{3}{=}33\pi/2$ and $\lambda_{2}{=}4\pi/5$, this second term is $\eta$ smaller than the first term given in Eq.(\ref{eq17}). For $\nu_p=\nu_0$, the polarization part reduces to $\left[|a_{1}|^{2}{+}|a_{-1}|^{2}\right]\left[|a_{2}|^{2}{+}|a_{{-}2}|^{2}\right]{=}1$ regardless the bright and dark excitons polarizations. If $N'$ instead of one dark exciton are involved, the corresponding term, proportional to $\eta^{N'}$, has a $|a_{2}|^{2N'}{+}|a_{{-}2}|^{2N'}$ factor which depends on the dark exciton polarization but still not on the probe photon polarization. To get a dependence on probe polarization and possibly test the linear polarization of the dark condensate, we need probe photons tuned on $\nu_p\neq\nu_0$.

The exciton shift $\langle \Psi_{N}|B_{p}V_{p}^{\dagger}|\Psi_{N}\rangle$ linear in Coulomb interaction corresponds to Fig. 2.a. Its dominant terms in $\eta$ are shown in Fig.~3. Again, a factor $N^2$ arises from the $N$ ways on each side to choose the dark exciton interacting with the bright exciton. The direct process, shown in Fig.~3a, cancels because the exciton states remain unchanged\cite{Phys_Rep}. We are left with the exchange Coulomb processes between one bright and one dark exciton shown in Fig.~3b. For $\nu_p=\nu_0$, they lead to
\begin{multline}\label{eq19}
    \langle \Psi_{N}|B_{p}V_{p}^{\dagger}|\Psi_{N}\rangle\simeq -N^{2}\xi^{exch}(_{0\;\;0}^{0\;\;0})
      \left[ \left|a_{1}a_{2}\right|^{2}+\left|a_{-1}a_{-2}\right|^{2}\right.\\ \left.+\left|a_{1}a_{2}\right|^{-2}+\left|a_{-1}a_{2}\right|^{2}\right]
(N-1)!F_{N-1}.
\end{multline}
Again, the bracket reduces to 1 regardless the bright and dark polarizations. As for $\langle \Psi_{N}|B_{p}B_{p}^{\dagger}|\Psi_{N}\rangle$, exchange Coulomb processes with $N'$ dark excitons depend on the dark exciton polarization but not on the probe polarization when $\nu_p=\nu_0$. As $\xi^{exch}(_{0\;\;0}^{0\;\;0})=-\xi_{d}(a_{X}/L)^{d}R_{X}$, where $R_{X}=e^{2}/2a_{X}$ is the 3D exciton Rydberg while $\xi_{3}=26\pi/3$ and $\xi_{2}=8\pi-315\pi^{3}/512>0$, first order exchange Coulomb processes with the dark condensate thus yield a blueshift of the excitonic absorption line
\begin{figure}\label{fig4}
\includegraphics[width=8.5cm]{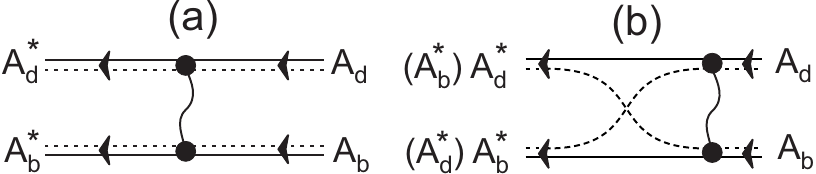}
\caption{Dominant terms of $\langle \Psi_{N}|B_{p}V_{p}^{\dagger}|\Psi_{N}\rangle$ in dark exciton density.}
\end{figure}

\vspace{-10 pt}

  \begin{equation}\label{eq20}
    \Delta\simeq\eta\,\xi_{d}R_{X}+\mathcal{O}(\eta^{2})
  \end{equation}
 independent of the probe polarization for photons tuned on the exciton ground state, this shift increasing with the dark exciton density.

The second term of Eq.(\ref{eq10}) also brings $\eta$ contributions to the shift when one among $N$ dark excitons is involved. This changes the numerical prefactor $\xi_{d}$ in the probe shift but not the structure of the result.
\begin{figure*}\label{fig5}
\includegraphics[width=18cm]{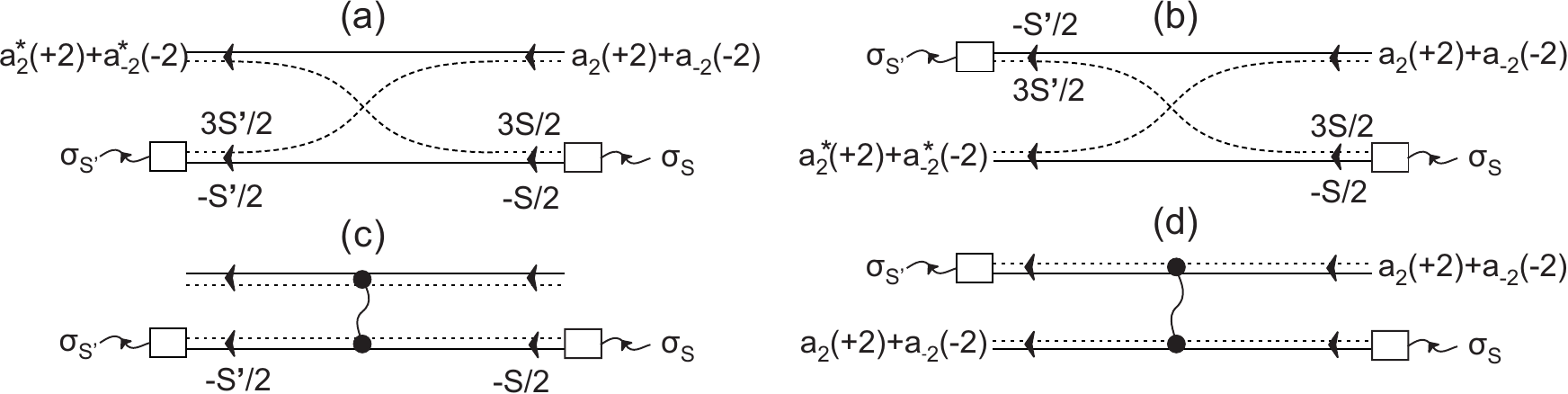}
\caption{Dominant contribution to $\Delta_{S',S}$ given in Eq. (18). $\Delta_{1,-1}\neq$0 would produce a Faraday oscillation while $\Delta_{1,1}\neq\Delta_{-1,-1}$ would produce a Faraday rotation.}
\end{figure*}

{\bf II. Dark excitons interacting with linearly polarized unabsorbed photons}.
Another effect driven by carrier exchanges is the interaction between excitons and an unabsorbed photon beam with linear polarization. When the excitons are bright, the problem is rather simple if their polarization is circular\cite{PRB_MC_OBM_2006}, let say $\sigma_+$. Through the virtual excitons to which the $\sigma_{+}$ part of a linearly polarized light is coupled, this $\sigma_{+}$ part  sees the exciton gas by electron exchange and by hole exchange, while no exchange exists for the $\sigma_{-}$ part. So, the $n_{\pm}$ indices are different and the polarization plane of the linearly polarized light rotates. The situation is far more complex when the bright excitons present in the sample have a linear polarization, let say along $\textbf{X}$. We may think that the $\sigma_{+}$ and $\sigma_{-}$ parts of the linearly polarized light have equal exchanges with the exciton gas; so, $n_{+}=n_{-}$ and nothing should happen. Yet, unabsorbed photons with polarization along $\textbf{X}$ or $\textbf{Y}$ should react differently to excitons with $\textbf{X}$
polarization. Since the linear polarization of the exciton gas does not differentiate $\sigma_{+}$ from $\sigma_{-}$ rotation, the polarization plane of the unabsorbed beam cannot rotate. It actually oscillates\cite{SSC_MC_OBM_2009}, with a polarization changing from linear to elliptical (or circular) and going back to linear again.

Let us reconsider this effect when the excitons present in the sample are dark. The major change comes from the fact that the two parts of the linearly polarized light see $S=2$ excitons by carrier exchange - hole for $\sigma_{+}$ and electron for $\sigma_{-}$ ; so, dark excitons should do nothing, regardless their polarization. To tackle this effect, we follow Ref. \cite{SSC_MC_OBM_2009} which leads to both, Faraday rotation and oscillation in the case of bright excitons.

Let $\alpha_{\pm1}^{\dagger}$ create an unabsorbed probe photon with energy $\omega_{p}$ and circular polarization $\sigma_{\pm}$.
In the absence of photon-semiconductor coupling, the two states

\vspace{-13 pt}

\begin{equation}\label{eq21}
    |\Psi_{N}^{(\pm1)}\rangle=\alpha_{\pm1}^{\dagger}|0\rangle\otimes|\Psi_N\rangle
    \end{equation}
    
    \vspace{-5 pt}
    
\noindent are degenerate in energy, $(H_{sc}+H_{ph}-\mathcal{\mathcal{E}}_{N}-\omega_{p}) |\Psi_{N}^{(\pm1)}\rangle=0$, where $H_{ph}$ is the photon hamiltonian.

In the presence of coupling, the system Hamiltonian reads $H=H_{sc}+H_{ph}+W$ with $W=U+U^{\dagger}$ and $U^{\dagger}=\sum\Omega_{j}B_{jS}^{\dagger}\alpha_{S}$. To get the $H$ eigenstates, $(H-\mathbf{E}_{N+1}^{'})|\Phi_{N+1}^{'}\rangle=0$, we proceed basically as we did for Eq.(\ref{eq6}) except that the  $|\Psi_{N}^{(\pm1)}\rangle$ subspace now is  degenerate. We introduce the projector $P_{\bot}$ over the subspace orthogonal to $ |\Psi_{N}^{(\pm1)}\rangle$ defined through
\begin{equation}\label{eq22}
   I=\frac{|\Psi_{N}^{(1)}\rangle\langle\Psi_{N}^{(1)}|}{\langle\Psi_{N}|\Psi_{N}\rangle}+
    \frac{|\Psi_{N}^{(-1)}\rangle\langle\Psi_{N}^{(-1)}|}{\langle\Psi_{N}|\Psi_{N}\rangle} +P_{\bot}.
\end{equation}

\noindent When inserted into the Schr\"{o}dinger equation for $|\Phi_{N+1}^{'}\rangle$ multiplied by $P_{\bot}$, we obtain an expression for $P_{\bot}|\Phi'_{N+1}\rangle$ similar to Eq.(5).
\noindent By using it into the Schr\"{o}dinger equation projected onto
$\langle\Psi_{N}^{(\pm1)}|$, we get 2 equations for $\langle\Psi_{N}^{(\pm1)}|\Phi_{N+1}^{'}\rangle$ which have a non-zero solution for\vspace*{-.3cm}
\begin{equation}\label{eq23}
    \begin{vmatrix}
 \mathcal{E}_{N}+\omega_{p}{+}\Delta_{{+}1,{+}1}{-}\mathbf{E}_{N{+}1}^{'} & \Delta_{{+}1,{-}1} \\
  \Delta_{{-}1,{+}1} & \mathcal{E}_{N}{+}\omega_{p}{+}\Delta_{{-}1,{-}1}{-}\mathbf{E}_{N{+}1}^{'}\\
 \end{vmatrix}=0.
\end{equation}
 \noindent The coupling between dark excitons and $\sigma_{\pm}$ photons reads, at lowest order in $U$, as
\begin{equation}\label{eq24}
    \Delta_{S',S}=\frac{\langle\Psi_{N}^{(S')}|U\big[\mathcal{E}_{N}+\omega_{p}-H_{sc}-H_{ph}\big]^{-1}U^{\dagger}|\Psi_{N}^{(S)}\rangle}{\langle\Psi_{N}|\Psi_{N}\rangle}.
\end{equation}

The state $U^{\dagger}|\Psi_{N}^{(S)}\rangle$ contains $N$ dark excitons plus one virtual exciton $j$ with spin $S$. Contribution to $\Delta_{S',S}$ linear in dark exciton density comes from interaction of this virtual exciton $(j,S)$ with one among $N$ dark excitons. These interactions can be pure carrier exchanges as in Figs.(4a,4b). They bring a contribution in $\eta$ multiplied by a polarisation factor given by
\begin{multline}\label{eq25}
\delta_{S',S}\bigg\{ \delta_{S,1}\left[|a_{2}|^{2}\lambda(_{j'\;\;j}^{0\;\;0})+|a_{-2}|^{2}\lambda(_{0\;\;j}^{j'\;\;0}) \right]\\+ \delta_{S,-1}\left[|a_{-2}|^{2}\lambda(_{j'\;\;j}^{0\;\;0})+|a_{2}|^{2}\lambda(_{0\;\;j}^{j'\;\;0}) \right]\bigg\}.
\end{multline}
This term is diagonal and identical for $S=\pm1$ if $|a_{2}|^{2}=|a_{{-}2}|^{2}$, i.e., when dark excitons are linearly polarized. The  $(j,S)$ exciton can also have direct Coulomb process with dark excitons, as in Figs.(4c,4d). Diagram (4c) leads to a contribution in $\delta_{S',S}\left[|a_{2}|^{2}+|a_{{-}2}|^{2}\right]$ again diagonal and independent of the dark polarization while the diagram of Fig.(4d) responsible for $\Delta_{1,{-}1}\neq0$ when bright excitons are present in the sample, reduces to zero when these excitons are dark. So, there is no interaction between dark and bright excitons possibly leading to a Faraday oscillation through $\Delta_{1,{-}1}\neq0$. When the dark condensate is linearly polarized, there also is no energy splitting between circularly polarized photons, $\Delta_{1,1}\neq\Delta_{{-}1,{-}1}$, possibly leading to a Faraday rotation: Absence of Faraday effect is an optical signature that  excitons condense into a linearly polarized dark state.

\textbf{As a conclusion}, we have shown that the presence of fully dark excitons can be optically revealed by a blueshift of the excitonic absorption line. This shift increases with the dark exciton density but depends on the probe photon polarization for photons \textit{not} tuned on the ground state exciton only. We also show that an unabsorbed photon beam with linear polarization is unaffected by the presence of a  dark condensate with linear polarization. 

\vspace*{.5cm}

\centerline{\large{Supplementary Material}}

\begin{center}
\textit{Here are the key equations of the many-body formalism for composite bosons that allow the derivation of Eqs.(10,18). More details can also be found in Ref.\cite{Phys_Rep,PRB_MC_OBM_2006}.}
\end{center}

\vspace*{.2cm}


\noindent (i) Many-body effects induced by carrier exchanges in the absence of Coulomb interaction follow from
(with $B_{m;S_m} \equiv B_{mS_m}$)
\begin{equation}\label{eq11}
\left[ B_{mS_{m}},B_{iS_{i}}^{\dagger} \right] =\delta_{mi}\delta_{S_{m}S_{i}}-D_{mS_{m},iS_{i}}\nonumber
\end{equation}

\begin{multline}\label{eq12}
  \left[ D_{mS_{m},iS_{i}},(B_{jS_{j}}^{\dagger})^{N} \right] = 
  N(B_{jS_{j}}^{\dagger})^{N-1}\\
  \sum_{nS_{n}}
  \left[
   \Lambda(_{mS_{m}\;\;iS_{i}}^{nS_{n}\;\;\;\; jS_{j}})
 \right. 
 \left.+ \Lambda(_{nS_{n}\;\;\;\;iS_{i}}^{mS_{m}\;\;\; jS_{j}})
 \right]
 B_{nS_{n}}^{\dagger}.\nonumber
\end{multline}
The Pauli scatterings split into an angular and a spin part according to
\begin{eqnarray}\label{eq13}
  \Lambda(_{mS_{m}\;\;\;iS_{i}}^{nS_{n}\;\;\;\; jS_{j}})
  = \lambda(_{m\;\;\;\;i}^{n\;\;\;\;j})\,
 \chi(_{S_{m}\;\;\;\;S_{i}}^{S_{n}\;\;\;\;S_{j}})
\nonumber
\end{eqnarray}
All the spin parts $\chi$ are equal to zero except (see Fig.1)
\begin{equation}\label{eq14}
\begin{array}{rclrcl}
  \chi(_{S\;\;\;\;S}^{S\;\;\;\;S})
 = 
 \chi(_{\epsilon\;\;\;\; -2\epsilon }^{-\epsilon\;\;\;\; 2\epsilon})
 =
 \chi(_{\epsilon\;\;\;\;\;\; \epsilon}^{2\epsilon\;\;\;\; 2\epsilon})
 =
   \chi(_{-2\epsilon\;\;\;\; \epsilon}^{\epsilon\;\;\;\; -2\epsilon})=1
   \end{array} \nonumber
\end{equation}
with $S=\pm 2$ or $\pm 1$, and $\epsilon=\pm1$.

\vspace*{.2cm}

\noindent (ii) Many-body effects induced by Coulomb interaction in the absence of carrier exchange follow from

\begin{equation}\label{eq15}
    \left[ H_{SC},B^{\dagger}_{iS_{i}}\right]=E_{i}B^{\dagger}_{iS_{i}}+V^{\dagger}_{iS_{i}}
\nonumber
\end{equation}

\begin{equation}\label{eq16}
    \left[ V^{\dagger}_{iS_{i}},(B^{\dagger}_{jS_{j}})^N\right]=N (B^{\dagger}_{jS_{j}})^{N-1}\sum_{mn}\xi(_{mi}^{nj})
   B^{\dagger}_{mS_{i}}B^{\dagger}_{nS_{j}}.\nonumber
\end{equation}


\end{document}